\documentclass[12pt]{article}

\usepackage{graphicx}
\begin{document}

\begin{center}
{\bf  Bouncing cosmology, $F(T)$ teleparallel gravity and entropy of apparent horizon} \\
\vspace{5mm} S. I. Kruglov
\footnote{E-mail: kruglov@rogers.com}
\underline{}
\vspace{3mm}

\textit{Department of Physics, University of Toronto, \\60 St. Georges St.,
Toronto, ON M5S 1A7, Canada\\
Canadian Quantum Research Center, \\
204-3002 32 Ave., Vernon, BC V1T 2L7, Canada} \\
\end{center}

\begin{abstract}
Two parameters scale factor leading to bouncing cosmology is considered. We show that at some model parameters we obtain the deceleration parameter $q_0\approx -0.535$ at the current epoch which is in agreement with the Planck data. The equation for the transition point when the universe expands from acceleration to deceleration phases is obtained. We find the equation for the function $F(T)$ within the teleparallel gravity with torsion field $T$ which provides bouncing cosmology. For some parameters of the model the function $F(T)$ was computed.
At the same time, in the framework of entropic cosmology, the associated entropy was obtained for particular model parameters. The equation of state for dark energy was obtained.
\end{abstract}

\section{Introduction}

The standard model of the universe evolution suffers the problem of the initial singularity which can be solved by using the bouncing cosmology \cite{Mukhanov,Mukhanov1}. Bouncing cosmology scenarios were investigated in various approaches \cite{Veneziano,Khoury,Biswas,Saridakis,Brandenberger,Saridakis1,Bojowald,Martin}. These scenarios were generalized by considering a periodic sequence of the universe contractions and expansions \cite{Novello}. At the same time $F(T)$ teleparallel gravity, which uses the Weitzenb\"{o}ck connection \cite{Weitzenb} (not the Levi-Civita connection), is of great interest because it can explain the acceleration of the universe (the dark energy). The field equations in the $F(T)$ gravity \cite{Bengochea,Linder} are the second order which is an advantage compared with $F(R)$ gravity possessing the fourth order equations. The space-time does not have the curvature but has the torsion $T$. In \cite{Jawad} the exponential $f(T)$ gravity ($F(T)=T+f(T)$) for describing cosmic inflation with a scalar field and different potentials was investigated.
The cosmological dark energy model with power law scale factor, incorporating the effect of viscosity, in the framework of generalized teleparallel gravity was studied in \cite{Jawad1}. In \cite{Jawad2} new holographic polytropic dark energy model within $f(T)$ gravity was considered. The pilgrim dark energy and generalized ghost dark energy models were investigated in \cite{Shahzad}. In \cite{Pervaiz} the cosmic evolution of an accelerating universe within the framework of the Einstein–Cartan–Sciama–Kibble theory, employing a flat, homogeneous and isotropic model, was examined. In Refs. \cite{Sharif,Sharif1,Sharif2,Sharif3,Sharif4,Sharif5,Sharif6,Sharif7,Sharif8,Sharif9}  the late time cosmic evolution was investigated in the models of gravities beyond Einstein's General Relativity.

The Hubble parameter $H$ is negative for the contracting phase before the bounce, and in the expanding phase after the bounce the Hubble parameter $H$ is positive. The continuity equation shows that at the bounce point $H = 0$.
Throughout the transition from contracting to expanding universe we have $\dot{H}>0$, but for the transition from expansion to contraction $\dot{H}<0$. We will start from the bouncing scale factor $a(t)$ possessing two parameters, and then derive the Hubble rate $H(t)$.
A scenario of bouncing cosmology in the early universe governed by $F(T)$ gravity and corresponding to an entropy of the apparent horizon will be considered.

In the following we have set $c=\hbar=1$.

\section{The model}

For the case of the Friedmann--Lema\^{i}tre--Robertson--Walker spatially flat universe the metric is given by
\begin{equation}
ds^2=-dt^2+a(t)^2(dx^2+dy^2+dz^2).
\label{1}
\end{equation}
Let us consider a bouncing scenario of universe with the scale factor
\begin{equation}
a(t) = a_B(1+\alpha t^2)^n,
\label{2}
\end{equation}
where $a_B$ is the scale factor at the bouncing point, and $\alpha$ is a positive parameter which describes the duration of the
bounce and $n$ is dimensionless parameter. This ansatz represents the nonsingular bouncing behavior. We will show that analytic solutions take place. The cosmic time $t$ range is ($-\infty,+\infty $) and $t = 0$ corresponds to the bouncing point. The case of $n=1/3$ was considered in \cite{Cai0,Cai} corresponding to the matter dominated contraction and expansion.  Other cases with $n=2/3$ and $n=1/2$ where studied in \cite{Mishra} and \cite{Sahoo}, correspondingly.
Introducing the dimensionless parameter $\bar{t}=\sqrt{\alpha}t$, Eq. (2) becomes $a(\bar{t}) = a_B(1+\bar{t}^2)^n$. The plot of $a(\bar{t})/a_B$ versus $\bar{t}$ is depicted in Fig. 1 for $n=1/3,1/2,2/3$.
\begin{figure}[h]
\includegraphics [height=2.0in,width=3.7in] {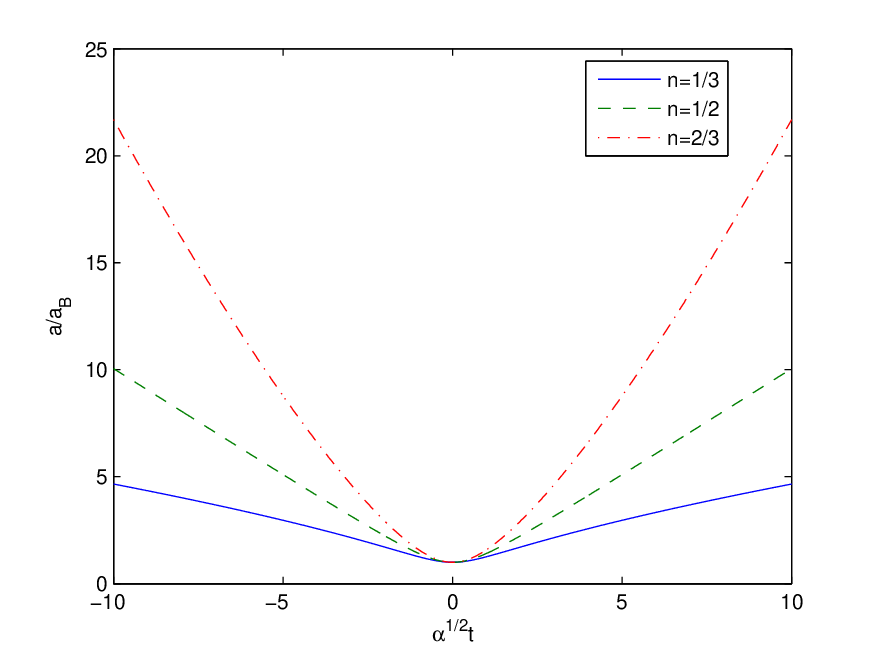}
\caption{\label{fig.1} The scale factor $a(\bar{t})/a_B$ vs. $\bar{t}$ for $n=1/3,~1/2,~2/3$. When parameter $n$ increases the scale factor also increases. The bouncing point corresponds to time $t=0$. When $t<0$ the scale factor is the decreasing function of the time and for $t>0$ $a(t)/a_B$ increases.}
\end{figure}
Figure 1 shows that when parameter $n$ increases the scale factor also increases. Using the definition of the Hubble parameter $H=\dot{a}/a$, one can directly derive
\begin{equation}
H(t) = \frac{2\alpha nt}{1+\alpha t^2}.
\label{3}
\end{equation}
By using the dimensionless Hubble parameter $\bar{H}=H/\sqrt{\alpha}$ Eq. (3) takes the form $\bar{H}=2n\bar{t}/(1+\bar{t}^2)$. The $\bar{H}$  versus $\bar{t}$ is plotted in Fig. 2 which shows the behavior of the reduced Hubble parameter $\bar{H}$ versus reduced time $\bar{t}$ for different parameters $n$.
\begin{figure}[h]
\includegraphics [height=2.0in,width=3.7in] {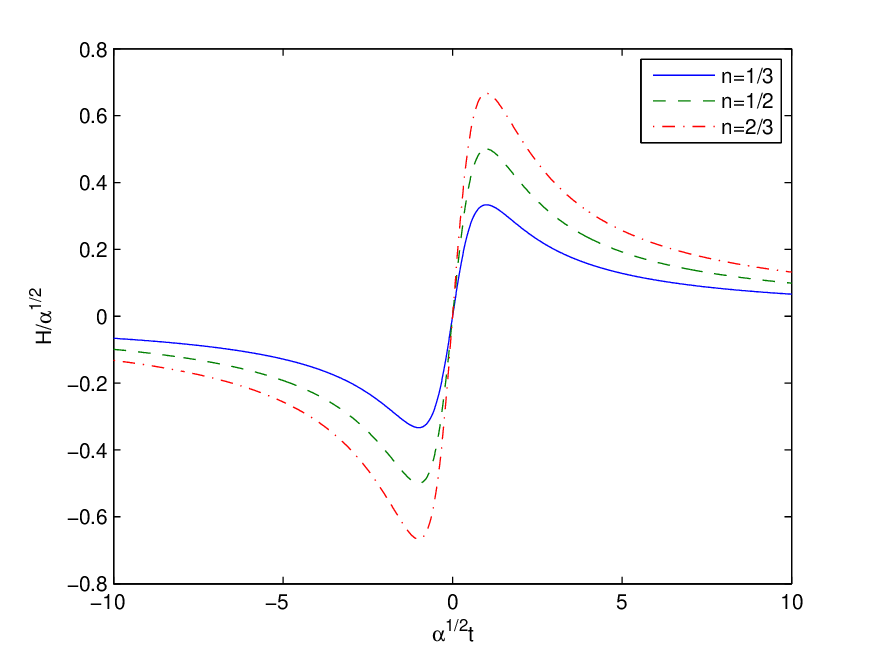}
\caption{\label{fig.2} The $\bar{H}$ vs. $\bar{t}$ at $n=1/3,~1/2,~2/3$. At $t>1/\sqrt{\alpha}$ and at $t<-1/\sqrt{\alpha}$ the reduced Hubble parameter $\bar{H}$ decreases. The duration of the bounce occurs at $1/\sqrt{\alpha}>t>-1/\sqrt{\alpha}$.}
\end{figure}
The duration of the bounce is given by the interval $1/\sqrt{\alpha}>t>-1/\sqrt{\alpha}$.

The deceleration parameter is defined by
\begin{equation}
q =-1- \frac{\dot{H}}{H^2}.
\label{4}
\end{equation}
When $q<0$ the acceleration phase takes place and at $q>0$ we have the deceleration phase. By virtue of Eqs. (2), (3) and (4) we obtain
\begin{equation}
q = \frac{(1-2n)\alpha t^2-1}{2n\alpha t^2}.
\label{5}
\end{equation}
The deceleration parameter $q$ versus reduced time $\bar{t}=\sqrt{\alpha}t$ at $n=1/3,~1/2,~2/3$ is plotted in Fig. 3.
\begin{figure}[h]
\includegraphics [height=2.0in,width=3.7in] {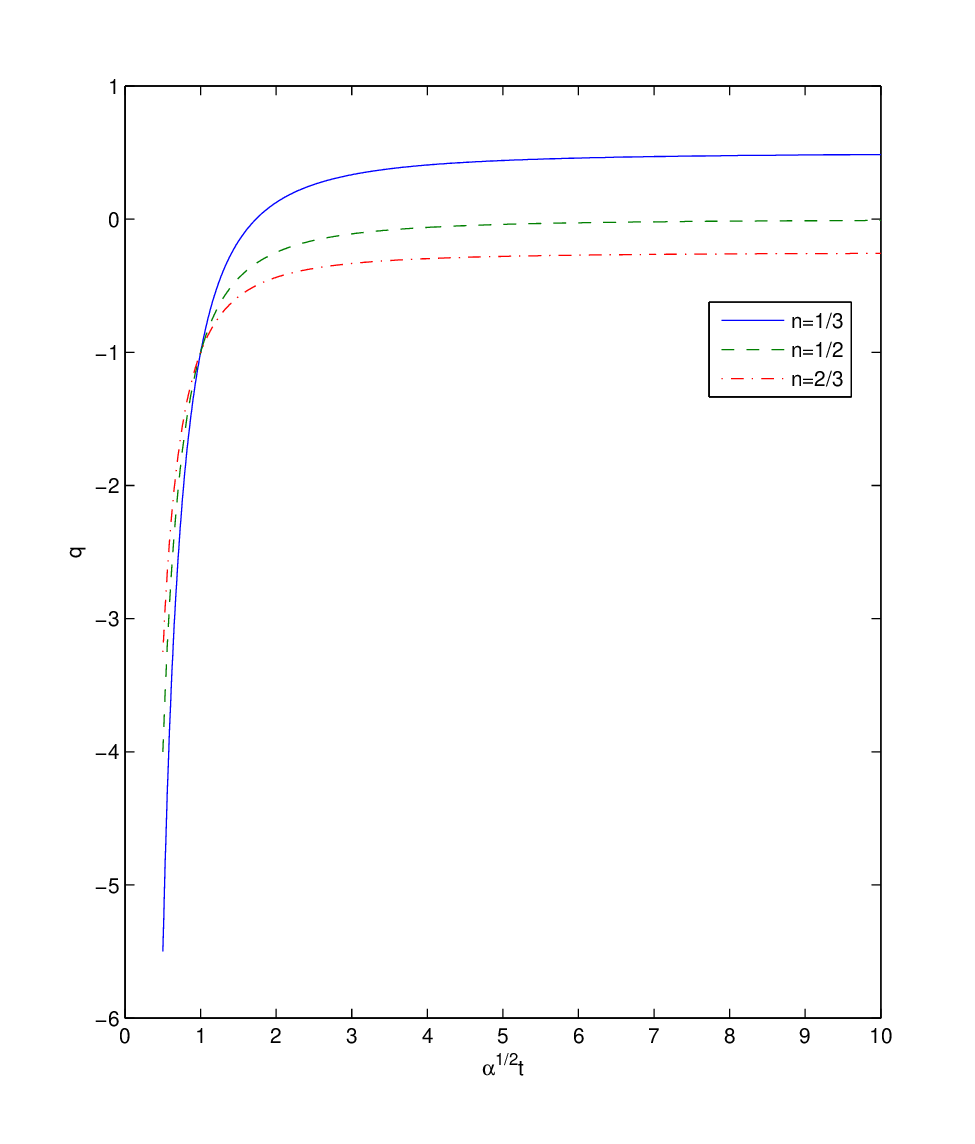}
\caption{\label{fig.3} The deceleration parameter $q$ vs. reduced time $\bar{t}=\sqrt{\alpha}t$ at $n=1/3,~1/2,~2/3$. Only acceleration phase takes place at $n>0.5$ and at $n<0.5$ we have two phases, the universe acceleration and deceleration. The point $q = 0$ corresponds to the transition from acceleration to deceleration phase.}
\end{figure}
 At $n>0.5$ we have only the acceleration phase corresponding to the eternal inflation. But at $n<0.5$ the two phases, the universe acceleration and deceleration take place. From Eq. (5) at $q = 0$ we obtain the equation for the transition time $t_{tr}=1/\sqrt{(1-2n)\alpha}$. Figure 3 illustrates this behaviour of the deceleration parameter. According to the Planck data \cite{Aghanim} the current value of the deceleration parameter is $q_0=-0.535$. From Eq. (5) we obtain the corresponding time
 \begin{equation}
t_0 = \frac{1}{\sqrt{(1-0.93n)\alpha}}.
\label{6}
\end{equation}
The value $t_0$ is less than the transition time $t_{tr}$ as should be. As an example, let $n=1/3$ for the matter bouncing scenario. Then $t_0\approx1.2/\sqrt{\alpha}$ and $t_{tr}\approx1.73/\sqrt{\alpha}$. Thus, after the bouncing point the universe accelerates till the present time when, according to Planck data, the deceleration parameter $q_0=-0.535$. Then the universe accelerates till the transition point $t_{tr}$ and then the universe decelerates till the bouncing point.

As a result, the model with two parameters considered allows us to have the scenario of bouncing cosmology with the deceleration parameter $q_0$ at current era which meets the Planck data.

\section{F(T) teleparallel gravity}

Within Teleparallel Equivalent to General Relativity (TEGR), instead of the curvature the torsion is introduced. In such theory of gravity the dynamics is similar to the dynamics of the General Relativity theory. Vierbein (tetrad) fields $e^A_{~\mu}$ in TEGR describe the geometry of space-time and define a torsion tensor. Tetrad fields are the source of gravity and there is the antisymmetric contribution to the Christoffel connection. The torsion tensor defines the torsion scalar $T$ which enters the gravitational action. In $F(T)$ gravity the Lagrangian density of TEGR is modified by using an arbitrary function of the torsion scalar. The space-time metric $g_{\mu\nu}$ is constructed from tetrads as
$g_{\mu\nu}=\eta _{AB}e^A_{~\mu}e^B_{~\nu}$, where $\eta _{AB}=\mbox{diag}(-1,+1,+1,+1)$ being the Minkowski metric in the local frame and indexes $A$ and $B$ label the orthonormal frame and $\mu$, $\nu$ label space-time coordinates. In the teleparallel gravity
the Weitzenb\"{o}ck connection is used, $\Gamma^\lambda_{~\mu\nu}=e_A^{~\lambda}\partial_\mu e^A_{~\nu}$. In the Weitzenb\"{o}ck connection the Riemann curvature tensor vanishes ($R_{\alpha\beta\mu\nu}=0$). The torsion tensor is defined as
 \begin{equation}
T^\rho_{~\mu\nu}=e^{~\rho}_A\left(\partial_\mu e^A_{~\nu}-\partial_\nu e^A_{~\mu}\right),
\label{7}
\end{equation}
and the superpotential tensor is
\begin{equation}
S_\rho^{~\mu\nu}=\frac{1}{2}\left(K_{~~\rho}^{\mu\nu}+\delta^\mu_\rho T_{~~\alpha}^{\alpha\nu}-\delta^\nu_\rho T_{~~\alpha}^{\alpha\mu}\right),
\label{8}
\end{equation}
where
\begin{equation}
S^{\mu\nu}_{~~\rho}=-\frac{1}{2}\left(T^{\mu\nu}_{~~\rho}-T^{\nu\mu}_{~~\rho}-T^{~\mu\nu}_\rho\right),
\label{9}
\end{equation}
is the contorsion tensor.
The torsion field $T$ is given by $T=S_\rho^{~\mu\nu}T^\rho_{~\mu\nu}$. For FLRW metric (1) $e^A_{~\mu}=\mbox{diag}(1,a,a,a)$ and
the torsion is $T=-6H^2$ \cite{Bengochea}.
Then from Eq. (3) we obtain the torsion
\begin{equation}
T =- \frac{24\alpha^2 n^2t^2}{(1+\alpha t^2)^2}.
\label{10}
\end{equation}
Equation (10) can be represented as the biquadratic equation $\alpha^2Tt^4+(2\alpha T+24\alpha^2n^2)t^2+T=0$ with the solution
\begin{equation}
t(T) =\pm \sqrt{-\frac{1}{\alpha}-\frac{12n^2}{T}+\frac{2n\sqrt{6\alpha(T+6\alpha n^2)}}{T\alpha}}.
\label{11}
\end{equation}
The solution (11) with the signs $+$ and $-$ correspond to the expansion and contraction  phases, correspondingly.
At $\alpha=3\sigma/2$ and $n=1/3$ one finds from Eq. (11) the result of Ref. \cite{Cai0}.

From $F(T)$ gravity equations we get the Friedmann equation \cite{Bengochea}
\begin{equation}
\frac{1}{6}\left[F(T)-2TF'(T)\right]|_{T=-6H^2}=\left(\frac{8\pi G}{3}\right)\rho,
\label{12}
\end{equation}
where $\rho$ is the matter density. By virtue of the continuity equation
\begin{equation}
\dot{\rho}=-3H(\rho+p),
\label{13}
\end{equation}
and utilizing Eq. (2) we obtain the density of the matter
\begin{equation}
\rho=\rho_B(1+\alpha t^2)^{-3n(1+w)},
\label{14}
\end{equation}
where $\rho_B$ is the matter density at the bouncing point and $w=p/\rho$ being the equation of state (EoS) for the matter and $p$ is the matter pressure. Making use of Eqs. (11) and (14) we represent Eq. (12) as follows:
\begin{equation}
F'(T)-\frac{F(T)}{2T}=-\frac{\rho_B}{TM_{Pl}^2}\left(\frac{2n}{T}\sqrt{6\alpha(T+6\alpha n^2)}-\frac{12\alpha n^2}{T}\right)^{-3n(1+w)},
\label{15}
\end{equation}
where $M_{Pl}$ is the reduced Planck mass, $M_{Pl}=1/\sqrt{8\pi G}$. The solution to Eq. (15) is given by
\begin{equation}
F(T)=-\frac{\rho_B\sqrt{T}}{2^CM_{Pl}^2}\int \frac{T^{C-3/2}dT}{(\sqrt{A(T+A)}-A)^C},
\label{16}
\end{equation}
where $A=6\alpha n^2$, $C=3n(1+w)$. The integral in Eq. (16) can be calculated analytically only for particular cases of the parameter $C$. Let us consider two cases, $C=0.5$ and $C=1.5$.

1. $C=3n(1+w)=0.5$. For the dust matter $w=0$, we have $n=1/6$. In this case the integral in Eq. (16) is expressed in the form of elementary functions and we have
\begin{equation}
F(T)=-\frac{\rho_B\sqrt{T}}{M_{Pl}^2}\left(\frac{1}{\sqrt{A}}\arctan\left(\sqrt{\frac{\sqrt{A(T+A)}-A}{2A}}\right)
-\sqrt{\frac{2}{\sqrt{A(T+A)}-A}}\right).
\label{17}
\end{equation}
Taking into account that $T=-6H^2<0$ and $i\arctan(ix)=-\mbox{tanh}^{-1}(x)$, we represent the real function $F(T)$ as
\begin{equation}
F(T)=\frac{\rho_B\sqrt{-T}}{M_{Pl}^2}\left(\frac{1}{\sqrt{A}}\mbox{tanh}^{-1}\left(\sqrt{\frac{A-\sqrt{A(T+A)}}{2A}}\right)
-\sqrt{\frac{2}{A-\sqrt{A(T+A)}}}\right),
\label{18}
\end{equation}
with $T=-6H^2$.

2. $C=1.5$. For the dust matter $w=0$, we have $n=0.5$. Then after integration we obtain from Eq. (16) the function
\begin{equation}
F(T)=\frac{\rho_B\sqrt{-T}}{2M_{Pl}^2}\left(\frac{4\left(2A-\sqrt{A(A+T)}\right)\sqrt{A-\sqrt{A(T+A)}}}{A\left(A-\sqrt{A(A+T)}\right)}\right).
\label{19}
\end{equation}
From Eq. (16), one can easily obtain the function $F(T)$ for other values of the parameter $C$.

Thus, we have showed that cosmology based on the scale factor with two parameters is equivalent to cosmology within $F(T)$ teleparallel gravity.
The function $F(T)$ depends only on parameter $C=3n(1+w)$.

\section{Entropic cosmology}

Let us consider the possibility of bouncing cosmology in the framework of apparent horizon thermodynamics. The radius of the apparent horizon ($R_h = a(t)r$) is $R_h=1/H$. The first law of apparent horizon thermodynamics reads
\begin{equation}
dE=-T_hdS_h+WdV_h,
\label{20}
\end{equation}
where $S_h$ is apparent horizon entropy, the work density is $W=(\rho-p)/2$ \cite{Hayward,Hayward1} and $E=(4\pi/3)\rho R_h^3$ is the total energy inside the space. The apparent horizon temperature is given by \cite{Cai1}
\begin{equation}
T_h=\frac{H}{2\pi}\left|1+\frac{\dot{H}}{2H^2}\right|.
\label{21}
\end{equation}
Making use of first law of apparent horizon thermodynamics (20), and utilizing Eqs. (21) and (13) we obtain \cite{Kruglov,Kruglov1}
\begin{equation}
\dot{S}_h=-\frac{8\pi^2 \dot{\rho}}{3H^4},
\label{22}
\end{equation}
where $\dot{S}_h=\partial S_h/\partial t$. Making use of Eq. (14) and integrating Eq. (22) we obtain the entropy
\begin{equation}
S_h=\frac{\pi^2(1+w)\rho_B}{\alpha^3n^3}\int \frac{(1+\alpha t^2)^{3-C}}{t^3}dt=\frac{\pi^2(1+w)\rho_B}{2\alpha^2n^3}B(1+\alpha t^2;4-C,-1),
\label{23}
\end{equation}
where $B(z;a,b)$ is incomplete beta function
\begin{equation}
B(z;a,b)=\int_0^z u^{a-1}(1-u)^{b-1}du.
\label{24}
\end{equation}
Equation (23) allows us to obtain the apparent horizon entropy associated with our model of bouncing cosmology for different parameters $\alpha$, $n$, and $w$. Let us consider some cases when the entropy (23) is expressed in the form of elementary functions.

1) $C=3n(1+w)=1$. As a particular case this is realized at the matter dominated epoch, $w=0$, $n=1/3$. In this case from Eqs. (23) and (24) we obtain the entropy
\begin{equation}
S_h=\frac{\pi^2(1+w)\rho_B}{2n^3
\alpha^2}\left(1+\alpha t^2-\frac{1}{\alpha t^2}+2\ln(\alpha t^2)\right).
\label{25}
\end{equation}
Making use of Eq. (11) equation (25) can be represented in terms of torsion $T$ or via relation $T=-6H^2$ through the Hubble rate. One can also express the entropy $S_h$ via Bekenstein--Hawking entropy $S_{BH}=\pi R_h^2/G=\pi/(GH^2)$. To have the positive entropy we find the solution
$S_h=0$ which leads to $\sqrt{\alpha} t\approx 0.883$. Thus, at $-0.883/\sqrt{\alpha}>t$ and $t>0.883/\sqrt{\alpha}$ the entropy is positive and possesses physical meaning. At $t>0.883/\sqrt{\alpha}$ the entropy is increasing function and at $-0.883/\sqrt{\alpha}>t$ the entropy is decreasing function that are the necessary requirements.

2) C=2. It is realized at $n=2/3$, $w=0$. Then from Eqs. (23) and (24) we obtain
\begin{equation}
S_h=\frac{\pi^2(1+w)\rho_B}{2n^3
\alpha^2}\left(-\frac{1}{\alpha t^2}+\ln(\alpha t^2)\right).
\label{26}
\end{equation}
We obtain the solution $S_h=0$, $\sqrt{\alpha} t=e^{W(1)/2}\approx 1.33$, where $W(x)$ is the Lambert function. At $-1.33/\sqrt{\alpha}>t$ and $t>1.33/\sqrt{\alpha}$ the entropy is positive. When $t>1.33/\sqrt{\alpha}$ the entropy is increasing function and at $-1.33/\sqrt{\alpha}>t$ the entropy is decreasing function. The plots of reduced entropy $\bar{S}_h=S_h\alpha^2n^3/((1+w)\rho_B)$ versus reduced time $\bar{t}=t\sqrt{\alpha}$ are depicted in Fig. 4.
\begin{figure}[h]
\includegraphics [height=2.0in,width=3.7in] {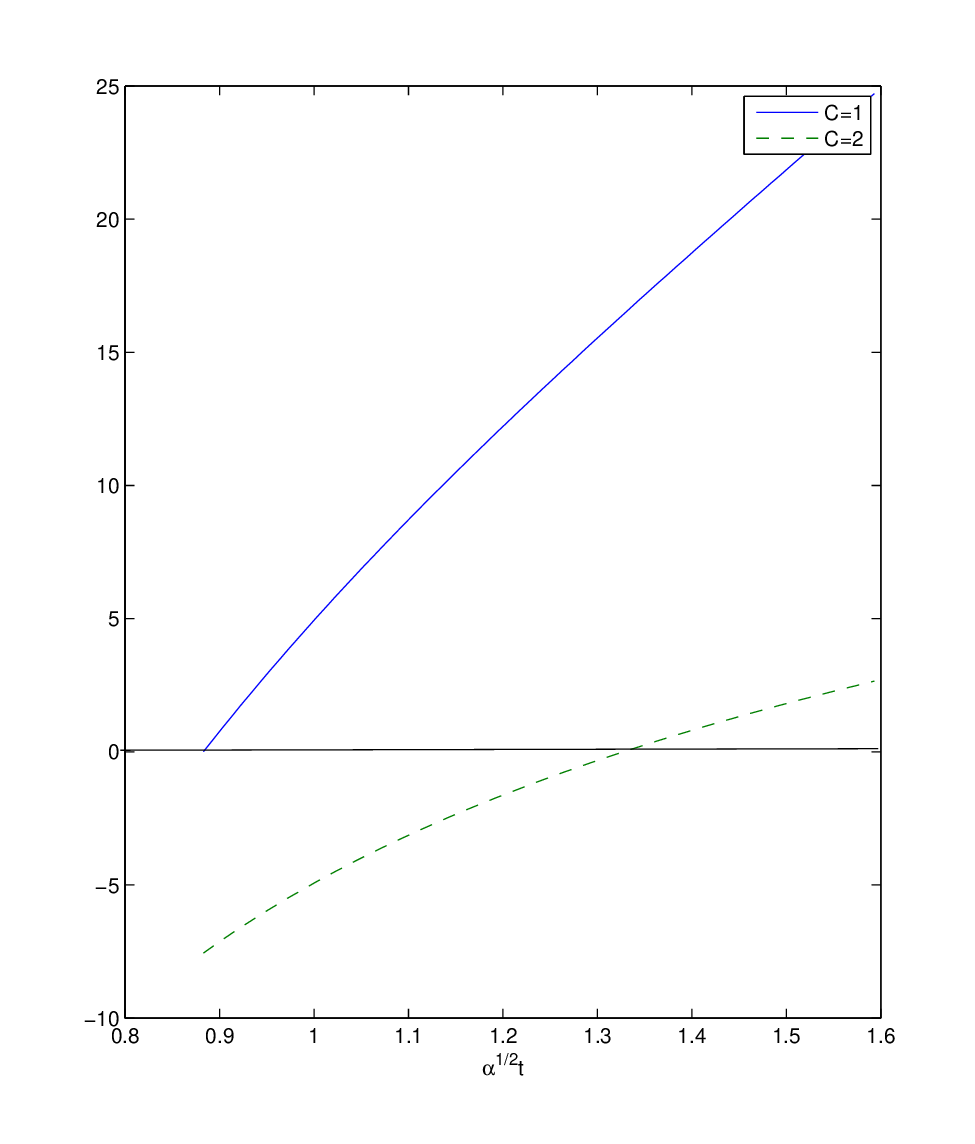}
\caption{\label{fig.4} Reduced entropy $\bar{S}_h=S_h\alpha^2n^3/((1+w)\rho_B)$ vs. reduced time $\bar{t}=t\sqrt{\alpha}$ for $C=1,~2$. For $C=2$ the time for $S_h=0$ is greater compared to the case $C=1$.}
\end{figure}
Fig. 4 shows that the entropy with parameter $C=1$ is growing with time more rapidly compared to the case $C=2$. From Eqs. (23) and (24) it is easy to get solutions for the entropy at $C=0.5$ and $C=1.5$. Entropy is formed after the bounce when $S_h=0$. Entropy grows in time because the radius the apparent horizon radius increase as well as the number of the degrees of freedom which, according to holographic principle, belongs to the boundary.

It was shown that cosmology based on the scale factor with two parameters leads to entropic cosmology with definite entropies computed. We have obtained the equation (23) for the entropy depending on three parameters $n$, $w$ and $\alpha$.

\subsection{Dark energy}

To study the dark energy, we consider the generalized Friedmann equation which is given by
\begin{equation}
H^2=\frac{8\pi G}{3}(\rho+\rho_D),
\label{27}
\end{equation}
where $\rho_D$ is the dark energy density and $\rho$ is the matter density. In general $\rho=\rho_m+\rho_r$ with $\rho_m$ being the density of the non-relativistic matter and $\rho_r$ is the density of the radiation. We assume that the dark energy obeys the conservation law (the continuity equation) $\dot{\rho_D}=-3H(\rho_D+p_D)$, where $p_D$ is the pressure corresponding to dark energy. Then differentiating on time Eq. (27) and utilization of the continuity equation for dark energy we obtain the generalized Friedmann equation for the pressure
\begin{equation}
H^2+\frac{2}{3}\dot{H}=-\frac{8\pi G}{3}(p+p_D).
\label{28}
\end{equation}
From Eqs. (27) and (28) we obtain the EoS for dark energy
\begin{equation}
w_D=\frac{p_D}{\rho_D}=-\frac{3H^2+2\dot{H}+8\pi Gp}{3H^2-8\pi G\rho}.
\label{29}
\end{equation}
Utilizing the Hubble parameter $H$, which is given by Eq. (3), and using Eq. (14) and $\dot{H}=2\alpha n(1-\alpha t^2)/(1+\alpha t^2)^2$ for the case of dust matter $p=0$, $w=0$, and $n=1/3$, we obtain from Eq. (29) the EoS for dark energy
\begin{equation}
w_D=\frac{\alpha}{6\pi G\rho_B(1+\alpha t^2)-\alpha^2 t^2}.
\label{30}
\end{equation}
At the time of the bounce $t=0$, we find the EoS for dark energy $w_D=\alpha/(6\pi G\rho_B)$.

Thus, we have computed the dark energy density $\rho_D$, pressure $p_D$ and EoS $w_D=p_D/\rho_D$ for the model considered.

\section{Summary}

We have introduced two parameters $\alpha$ and $n$ in the scale factor which leads to bouncing cosmology. It was shown that at some model parameters the deceleration parameter $q_0\approx -0.535$ at the current epoch takes place which is in agreement with the Planck data. We have obtained the equation for the transition point when the universe expands from acceleration to deceleration phase. The equation for the function $F(T)$ within the teleparallel gravity with torsion field $T$ was found providing bouncing cosmology. In the framework of entropic cosmology the entropy was obtained for some particular cases. In the scenario suggested the universe accelerates from the bounce point till the current time $t_0 = 1/\sqrt{(1-0.93n)\alpha}$ with the deceleration parameter $q_0=-0.535$ and then till the transition point $t_{tr} = 1/\sqrt{(1-2n)\alpha}$ ($n<0.5$) and then the universe decelerates till the bounce point. The equation of state for dark energy was obtained.
Thus, the bouncing cosmology takes place in the scenario considered.

\end{document}